\documentclass[aps,twocolumn]{revtex4}

\renewcommand{\sinh}{\mbox{sinh\,}}

\def\f_0{{f}}

\def\bR{{\bf R}}

\def\Im{\mathop{\rm Im}\nolimits}

\def\wt{\widetilde}
\def\ovl{\overline}

\def\interior#1{\setbox1=\hbox{$#1$}\rlap{$#1$}\kern0.4\wd1\raise1.1\ht1%
\hbox{$\scriptstyle \circ$}}

\def\boxit#1#2{\setbox1=\hbox{\kern#1{#2}\kern#1}%
\dimen1=\ht1 \advance \dimen1 by #1 \dimen2=\dp1 \advance \dimen2 by
#1
\setbox1=\hbox{\vrule height\dimen1 depth\dimen2\box1\vrule}%
\setbox1=\vbox{\hrule\box1\hrule}%
\advance \dimen1 by .4pt \ht1=\dimen1 \advance \dimen2 by .4pt
\dp1=\dimen2 \box1\relax}
\def\endprf{\raise .5ex\hbox{\boxit{2pt}{\ }}}
\def\w{w}

\def\half{{\scriptstyle{1 \over 2}}}

\def\ifundefined#1{\expandafter\ifx\csname#1\endcsname\relax}

\def\beq{\begin{equation}}
\def\endq{\end{equation}}
\def\beqa{\begin{eqnarray}}
\def\endqa{\end{eqnarray}}

\def\coupl{{\gamma}}

\begin{document}

\title{Lifetime of a massive particle in a de Sitter universe}

\author{Jacques Bros$^1$, Henri Epstein$^2$ and Ugo Moschella$^3$}
\affiliation{$^1$Service de Physique th\'eorique - CEA. Saclay.
91191 Gif-sur Yvette.\\$^2$Institut des Hautes \'Etudes
Scientifiques, 91440 Bures-sur-Yvette.\\$^3$Universit\`a
dell'Insubria, Como and INFN Milano}

\begin{abstract}
We study particle decay in  de Sitter space-time as given by first
order perturbation theory in an interacting quantum field theory.
We show that for fields with masses
above a critical mass $m_c$ there is no such thing as particle
stability, so that decays forbidden in flat space-time do occur
there. The lifetime of such a particle also turns out to be
independent of its velocity when that lifetime is comparable with de
Sitter radius. Particles with lower mass are even stranger:
The masses of their decay products must obey quantification rules,
and their lifetime is zero.
\end{abstract}

\maketitle Some important progress in the astronomical observations
of the last ten years \cite{Ri,Per} have led  to the surprising
conclusion that the recent universe is dominated by a "dark" exotic form
of energy density that acts repulsively at
large scales.
The simplest and best known candidate for the "dark energy"
is the cosmological constant, and the de Sitter geometry, which is the homogeneous and
isotropic solution of the cosmological Einstein equations in vacuo, appears to take the double role of reference
geometry of the universe, namely the geometry of spacetime deprived
of its matter and radiation content and of the  geometry that the
universe will approach asymptotically.

One might think that the presence of a cosmological constant, while
having a huge impact on our understanding of the universe as a
whole, would not influence microphysics in its quantum aspects. This
is also the viewpoint taken in the context of inflationary models
\cite{Linde}, where the effective cosmological constant is many
orders of magnitude larger than the one observed today. However this
conclusion may have to be reassessed. Indeed, in presence of a
cosmological constant, however small, it is the notion of elementary
particle itself which has to be reconsidered, since the usual
asymptotic theory is based on concepts which refer closely to
Minkowski spacetime and to its Fourier representation, and do not
apply to the de Sitter universe which is not asymptotically flat; in fact a true asymptotic theory does not exist
at present for de Sitter space. A possible basic approach is perturbation theory; unfortunately,
calculations of perturbative amplitudes which in the Minkowskian
case would be simple or even trivial become rapidly prohibitive or
impossible in the de Sitter case: this in spite of the fact that one
is dealing with a maximally symmetric manifold.

In this letter we have tackled one such calculation, namely that of
the mean lifetime of dS unstable scalar particles. The
results exhibit significant differences compared to the Minkowski case
and decay processes which are normally forbidden become
possible (as exhibited by Nachtmann \cite{nachtmann} in a special case)
and vice versa, processes that are normally possible are now
forbidden. The maximal symmetry  of the dS universe
allows for the introduction of a global mass operator, one of the two  Casimir operators of the dS group ${SO}(1,d)$
(see e.g. \cite{gursey});
this quantity is conserved for dS invariant field theories and it still makes sense to
follow Wigner \cite{Wigner} in associating a particle with a unitary
irreducible representation of the dS group
labeled by a mass parameter, as we do here.
However, in contrast with the Poincar\'e group,
the tensor product of two unitary irreducible representations
of masses $m_1$ and $m_2$ decomposes into a direct
integral of representations whose masses $m$
{\sl do not satisfy the "subadditivity condition" }
$m\geq m_1 + m_2$: all representations of mass larger than a certain
critical value (principal series) appear in the decomposition.
This fact was shown in \cite{Verdiyev} for the
two-dimensional case and will be established here in general. This
means that the de Sitter symmetry does not prevent a particle with
mass in the principal series from decaying into e.g. pairs of
heavier particles. This phenomenon also implies that there can be
nothing like a mass gap in that range. This is a major obstruction
to attempts at constructing a de Sitter S-matrix; the Minkowskian
asymptotic theory makes essential use of an isolated point  in the
spectrum of the mass operator, and this will generally not occur in
the de Sitter case. We
will also show that the tensor product of two representations of
sufficiently small mass below the critical value (complementary
series) contains an additional finite sum of discrete terms in the
complementary series itself (at most one term in dimension four).
This implies a form of particle stability,  but the new phenomenon
is that a particle of this kind cannot disintegrate unless the
masses of the decay products have certain quantized values.
Stability for the same range of masses has also been recently
found  \cite{skenderis} in a completely different context.

We will resort to first order perturbation theory in our
calculations. These are made trivial in the Minkowski case by the
use of momentum-space, but this is not so in the de Sitter space.
We restrict at first our
attention to the principal series and start by deriving a new general
formula expressing the decay probability of a particle which applies
to both Minkowski and de Sitter space-times in dimension $d$, where
the de Sitter manifold is identified with the hyperboloid $\{x \in
\bR^{d+1}\ :\ x^2 =x_0^2-x_1^2-\ldots-x_d^2= -R^2\}$  in the
$(d+1)$-dimensional Minkowski space. Differences will come in later.

A Klein-Gordon  neutral scalar field $\phi$ with mass $m \ge 0$ is
characterized by its two-point vacuum expectation value
$\w_{m}(x,\ y) = (\Omega,\ \phi(x)\,\phi(y)\Omega)$
which is uniquely specified up to a constant factor by
requiring  invariance, locality, and a suitable spectral condition
(having a thermodynamical physical interpretation
in the dS case \cite{bgm,Gibbons}). $\w_{m}$ allows for the
reconstruction of the Fock space of the theory and of a
representation of the invariance group whose
restriction to the one-particle subspace
is irreducible and labeled by  $m$. In the dS case  $m$ can
be related to a dimensionless parameter $\nu$ as follows
\begin{eqnarray} m^2R^2 &=& \left ( {d-1 \over 2} \right )^2 + \nu^2
\label{w.2}.\end{eqnarray} The range $m \geq  m_c = (d-1)/2R $ (i.e. $\nu$ real)
corresponds to the principal series  while $0 \le m
< m_c$ corresponds to the {complementary series} ($\nu$ imaginary).
These restrictions ensure that $\w_{m}$ is positive definite and
therefore a quantum theoretical interpretation is available. Consider now an interaction
\[
\int \coupl\,g(x)\,{\cal L}(x)\,dx,\ \ \ {\cal L}(x) =\
:\phi_0(x)\phi_1(x)^{q_1}\ldots\phi_N(x)^{q_N}: \label{1.3.3}\]
between  $1+N$  independent scalar fields
$\phi_0,\phi_1,\ \ldots,\ \phi_N$ with masses $ m_0,\ m_1,
\ldots,\ m_N$; self-interactions ${\cal L}(x) =\ :\phi_0(x)^{n}:$ are a special case
of this coupling. The spacetime dependent "switching-on factor"  $g(x)$
is there to take care of the infrared divergence of the integrals and amounts to putting the system in a box and allowing for a finite
time duration of the interaction. In the end however $g$ should be made to tend to 1 everywhere (adiabatic limit).

Let $\Psi = \int dx\, \f_0(x)\,\phi_0(x)\Omega$ be a (normalized)
one-particle state created by $\phi_0$ from the vacuum; $\f_0(x)$ encodes the physical details
about the quantum state of the unstable particle whose
disintegration we aim to study. At first order in perturbation theory  Wick's theorem
gives the transition  probability $\Gamma_{1_0;q_1,\ldots,q_N}$ from $\Psi$ to any possible state
containing $q_1+q_2 + \ldots + q_N$ particles created respectively by the fields $\phi_1,\phi_2,\ldots \phi_N$:
\begin{eqnarray}
&& \Gamma = \coupl^2\int dxdudvdy \ovl{\f_0(x)}\,\f_0(y)\,g(u)\,g(v)K(x,u,v,y), \label{gamma}\\
&&  K(x,u,v,y) =  \w_{m_0}(x,u)\, \prod_{j=1}^N q_j!\, \w_{m_j}(u,
v)^{q_j}\,\w_{m_0}(v, y).\nonumber \end{eqnarray}
The four-point kernel $K$ admits an easy graphical interpretation in $x$-space
as a bubble with two external legs.
The product of two-point functions (the bubble) may be replaced by
its {\em K\"all\'en-Lehmann type representation}:
\begin{equation}
\prod_{j=1}^N  \w_{m_j}(u,
v)^{q_j} = \int da^2 \rho(a^2;\ m_1,\ldots,\
m_n)\,\w_a(u, v). \label{1.3.1}\end{equation}
It is far from obvious that such representation exists; besides,
calculating $\rho$ is easy only in the
Minkowski case for $n=2$ (see below); difficult or impossible in other
cases. While the amplitude (\ref{gamma}) is infrared
divergent in the limit $g \to 1$, it is possible to
replace just one of the $g$'s  by 1 in the integral.
Then one integration can be done by means of the  {\it
projector identity} ($dy$ denotes either the Poincar\'e or the dS invariant measure):
\[
\int \w_m(x, y)\,\w_{m'}(y,\ z)\,dy = C(m)\delta(m^2-m^{\prime 2})
\w_m(x,z). \label{1.3}\]
Here the global structure of either the Minkowski or the dS spacetime enters crucially.
For dS theories this identity holds only for the principal series; $C(m) = 2\pi$ (Minkowski); $
 C(\nu) = 2\pi |\coth(\pi\nu)|$  (dS). There follows a general formula for the transition probability:
\begin{eqnarray} \Gamma_{1_0;q_1,\ldots,q_N}   &=&
{\coupl^2\,C(m_0)\,\int g(x)\,|F(x)|^2\,dx \over \int
\ovl{\f_0(x)}\w_{m_0}(x,\ y)\,\f_0(y)\,dx\,dy}
\nonumber\\
& \times & \left(\prod_{j=1}^N
q_j!\right)\,\rho(m_0^2;m_1,\ldots,m_N).\;\;\;\;
\label{1.17}\end{eqnarray} Here  $F(x) = \int \w_{m_0}(x,\
y)\,\f_0(y)\,dy$ is the wavefunction associated to $f$; the denominator is the squared norm of $\Psi$
no longer assumed to be one. This formula has an
interesting simple structure: the first factor does not depend on
the number or nature of the decay particles but only on the
wavefunction of the incoming unstable particle. The infrared problem
is contained in this factor and has to be overcome when letting the
remaining $g(x)$ tend to 1. The second factor is just
the relevant K\"all\'en-Lehmann weight times the right combinatorial
factor.

We now focus on the decay of a particle of mass $m_0$
into two  particles of mass $m_1$ and
briefly discuss  the well-known Minkowski case first.
The weight $\rho(m_0^2;m_1,m_1)$ can be computed by
Fourier transforming $\w^2_{m_1}(x,y)$:
\begin{equation}
\rho = \frac{\left( {m_0^2
-4m_1^2} \right )^{d-3 \over 2} }{\left(4\pi \right )^{d-3 \over 2} 2^d\,\pi m_0 \,\Gamma(\frac{d-1}{2})} \  \theta(m_0^2 -4m_1^2).
\label{m.2}\end{equation}
The appearance of Heaviside's $\theta$ function forbids the decay of a particle
into two that are globally heavier. This is a familiar consequence
of the Poincar\'e invariance of the theory. As for the
adiabatic limit, the common choice is to choose $g(x)$ as the
characteristic function of some time interval $T$.
It is then found
that the transition probability (\ref{1.17}) is proportional to $T$
and thus diverges when $T\rightarrow \infty$. Fermi's golden rule
tells us that the transition probability per unit time (see e.g.
\cite{Veltman}) has a finite limit
\begin{eqnarray}
\lefteqn{\frac 1 \tau=\lim_{T\to\infty}
\frac{\Gamma_{1_0;2_1}}{T}=2\rho(m_0^2;\ m_1,\ m_1) \times }
\nonumber\\
&&\;\;\;\;{(2\pi)\coupl^2\,\int (2p^0)^{-1}\,|\wt \f_0(p)|^2\,
\delta(p^2-m_0^2)\theta(p^0)\,dp \over \int |\wt
\f_0(p)|^2\,\delta(p^2-m_0^2)\theta(p^0)\,dp }\;\;\;\;
\label{m.1}\end{eqnarray}
$\tilde f (p)$ is the
Fourier transform of the wavepacket $f(x)$.
The crucial factor $(2p^0)^{-1}$ in the numerator controls the dependence of the result (\ref{m.1}) on the
wavepacket $\f_0$. In particular, the decay rate of a particle at rest in our frame can be obtained by letting $|\wt \f_0(p)|^2$
tend to $\delta(\vec{p})$.
In this limit 
the factor $(2p^0)^{-1}$  becomes $(2m_0)^{-1}$ and one gets:
\begin{eqnarray}
{\frac{1}{\tau_0} = {2^{1-d}\coupl^2 \over
\Gamma(\frac{d-1}{2})}\,\frac{1}{m_0^2}\,\left ( {m_0^2-4m_1^2 \over
4\pi} \right )^{\frac{d-3}2}\,\theta(m_0^2 -4m_1^2).  \label{m.4}} \label{tv}\end{eqnarray}
Had we chosen the wavepacket of a particle with sharp momentum $\vec p$,
we would have obtained an extra Lorentz factor and the
lifetime would be longer:
\begin{equation} \tau(\vec v) = {\tau_0}{{(1-{v^2}/{c^2})}^{-\frac 12}}, \ \ \ \ \vec v = c\vec p/p^0. \end{equation}
In the de Sitter case we use the dimensionless
parameter $\nu$ (see Eq. \ref{w.2}) to label the two-point
functions; they are proportional to Legendre functions of
the first kind:
\begin{eqnarray}
&&{ \w_\nu(z,z')\label{wig1} =\w_\nu(\zeta)=\frac{ \Gamma\left(\frac{d-1}{2} +i
\nu\right)\Gamma\left( \frac{d-1}{2} -i
\nu\right)}{2(2\pi)^{\frac{d}{2}} R^{d-2}} \times} \cr && \ \ \ \ \
\ \ \ \ \ \ \ \ \ \ \ \ \ \ \
\times{\;(\zeta^2-1)^{-\frac{d-2}{4}}\,
P^{-\frac{d-2}{2}}_{-\frac{1}{2} + i\nu}(\zeta)};
\label{legendre}
\end{eqnarray}
$z,z'$ belong to suitable tubular domains of the complex de Sitter spacetime;
 $\zeta = z\cdot z'/R^2$ in the ambient spacetime sense
(see \cite{bgm} for details).
For $d=2$ the positivity of $\rho(\nu^2;\nu,\ldots,\nu)$  has already
been established  \cite{nachtmann} for the disintegration
into an odd number of particles having  the same mass as the unstable particle.
Here we need to actually  compute the K\"all\'en-Lehmann weight
 i.e. to obtain $\rho(\kappa^2; \nu,\nu) \equiv \rho_\nu(\kappa)$ such that
\begin{equation}
\w^2_{\nu}(\zeta) = \int_0^{\infty} d\kappa^2 \rho_\nu(\kappa)\w_\kappa(\zeta) = \int_{-\infty}^{\infty} \kappa d\kappa
\rho_\nu(\kappa)\w_\kappa(\zeta). \label{klds}\end{equation}
Due to (\ref{legendre}), the generalized
Mehler-Fock theorem \cite{Magnus} tells us that $\rho_\nu(\kappa)$ can be expressed as the integral
\begin{eqnarray}
\lefteqn{\rho_\nu(\kappa) = \frac{ \left(\Gamma\left(\frac{d-1}{2}
+i \nu\right) \Gamma\left(\frac{d-1}{2} -i \nu\right) \right)^2
\sinh \pi \kappa } {2(2\pi)^{1+\frac{d}{2}} R^{d-2}}\times}  \cr &&
\!\!\!\!\!\!\times \int_1^\infty P^{-\frac{d-2}{2}}_{-\frac{1}{2} +
i \kappa} (x)\, {[P^{-\frac{d-2}{2}}_{-\frac{1}{2} +
i\nu}(x)]^2}{(x^2-1)^{-\frac{d-2}4}} \ dx \label{mf}
\end{eqnarray}
which is well defined for masses such that $|\Im\nu|<
\frac{d-1}{4}$; this includes the principal series and a portion of
the complementary series. Computing (\ref{mf}) involves only trigonometric functions for odd $d$
\cite{pasquier}. Remarkably,  Mellin
transform techniques \cite{Marichev} allow the computation for
any dimension $d$ (details will be published elsewhere):
\begin{eqnarray}
\rho_\nu(\kappa) &=& \frac{R^{2-d} {\sinh\pi\kappa} }{
(4\pi)^{\frac{d+2}{2}}\sqrt\pi \Gamma(\frac{d-1}2)} \frac{
\Gamma\left(\frac{d-1}4+ \frac{i\kappa}{2}\right)
\Gamma\left(\frac{d-1}4-
\frac{i\kappa}{2}\right)}{\Gamma\left(\frac{d+1}{4}
+\frac{i\kappa}{2}\right)\Gamma\left(\frac{d+1}{4}
-\frac{i\kappa}{2}\right)}\cr && \times
\prod_{\epsilon,\epsilon'=\pm}
\Gamma\left(\frac{d-1}4+{i\epsilon\nu}+
\frac{i\epsilon'\kappa}{2}\right) \label{weight}
\end{eqnarray}
Contrary to (\ref{m.2}),  the weight $\rho$ never vanishes.
This means that for $m> m_{c}$ decay processes into heavier particles
are always possible and thus, in that range of masses, one is not allowed to draw
conclusions about the stability of a certain particle just from its
being the lightest in a hierarchy. The Minkowskian  result
(\ref{m.2}) is however recovered in the limit of zero curvature ($R\to \infty$) that is
achieved by setting $\kappa=m_0R$ and $\nu=m_1R$.
Lowest order corrections to the flat case give: \beqa
&&\!\!\!\!{{R^2\rho_{m_1R}(m_0 R)} \sim {|\Delta m|^{d-3\over 2}
\over 2^d \pi^{d-1\over 2} \Gamma \left ({d-1 \over 2} \right ) m_0}
\left ( {m_0+2m_1\over 4} \right )^{d-3\over 2} }\times\cr &&
\times\left( 1 + \frac{A}{R^2}\right) \left[\theta(\Delta m)
+e^{-|\Delta m| R}\theta(-\Delta m)\right], \label{l.5}\endqa
$\Delta m = m_0-2m_1$. The lack of particle stability
($\Delta m < 0$) is exponentially small in $R$. If $\Delta m > 0$
there is a correction to the flat case of the order of the
cosmological constant $\Lambda = \frac{(d-1)(d-2)}{2R^{2}}$.
In the four dimensional case
\begin{equation}
A= \frac{17}{64 \left(m_1+\frac{m_0}{2}\right)^2}-\frac{107}{24
m_0^2}+\frac{17}{64 \left(\frac{m_0}{2}-m_1\right)^2}\end{equation}

All these effects are of course extremely small with the current
value of the cosmological constant. What about particle physics at
inflation? At that epoch $m R \sim m \times 10^{-15} {\rm GeV}^{-1}
\ll \frac{3}{4} $ for every particle of reasonable mass. Our results
should therefore be extended to the remaining portion of the
complementary series $\frac{d-1}4< |\Im \nu| <  \frac{d-1}2$ where all scalar
particles lie at the inflation era (but: there is no complementary
series in the Fermionic case).
By analytic continuation of (\ref{weight}) in $\nu$,
\begin{eqnarray}
\w_\nu^2 = \int_{-\infty}^{\infty}\kappa\,d\kappa\
\rho_\nu(\kappa)\w_\kappa\,\
+ \sum_{n=0}^{N-1} A_n(\nu)\,\w_{i(\mu +2i\nu+2n)}\nonumber  \label{h.14}\\
\begin{array}{lll} A_n(\nu) & = & {8\pi (-1)^n \over n!2^d
\pi^{1+d\over 2}R^{d-2}\Gamma(\mu)} \frac{\Gamma(\mu
+2i\nu+n)\Gamma(-2i\nu-n)}{ \Gamma(\mu+2i\nu+2n)\Gamma(-\mu-2i\nu
-2n)} \\ & &\hfill \times \frac{\Gamma(\mu+n)\Gamma(-i\nu-n)
\Gamma(\mu+i\nu+n)}{ \Gamma( -i\nu-n+\half) \Gamma(\mu+i\nu+n
+\half)}\ \nonumber \label{h.15}
\end{array}
\end{eqnarray}
where $\mu=(d-1)/2$. The number of discrete terms is the largest $N$
satisfying $N < 1 + |\Im \nu|-\mu/2$, or 0 if this is negative. A
particle of the complementary series with parameter $\kappa=i\beta$
can only decay into two particles with parameter $\nu = \frac{i}{2}(
|\beta|+ \mu +2n)$, where $n$ is any integer such that $0 \le 2n <
\mu-|\beta|$, and the decay is instantaneous. A particle with mass
$m \ll m_c$ can only decay into two particles of mass  $m_1 \sim
m/\sqrt{2}$. Even if the geometry of the universe at inflation was not exactly de
Sitterian, this example indicates that quantum field theoretical
arguments concerning particle physics at inflation might need
revision.

In trying to interpret the above results one can wonder
whether they might be due to the thermodynamical properties
\cite{bgm,Gibbons} of the fundamental state we have been using.
We have tested this possibility against a similar computation
in flat thermal field theory that however does not exhibit
this phenomenon in two-particle decays.
Another issue has to do with energy conservation and the relation
mass/energy. dS invariant field theories
admit ten conserved quantities (in $d=4$).  The identification
of a conserved energy  among these quantities
has proven to be useful in classical field theory \cite{abbott}.
The same quantity remains exactly
conserved also at the quantum level although it becomes an operator whose spectrum
is not positive \cite{bgm} even when restricted to the region where the corresponding classical expression is positive \cite{abbott}; the thermodynamical properties of dS fields arise precisely in this restriction  \cite{bgm}.
Energy is conserved also in the  decay processes that violate mass subadditivity, once the adiabatic limit has been performed.
The breakdown of the subadditivity property of
masses in dS spacetime just reflects the nonexistence of
an Abelian translation group and thereby
of a linear energy-momentum space.

We now consider the adiabatic limit problem and its meaning in the de Sitter
context, in the case when all particles are in the principal series.
A first complication is the existence of several  choices
of cosmic time, having different physical implications and the
result might depend on one's preferred choice. In the closed model,
the cosmic time $t$ is related to the ambient space coordinates as
follows: $x^0 = R\, \sinh (t/R)$. In strict analogy to the Minkowski
case, $g(x)$ can be chosen as the indicator function of some cosmic
time interval $T$, say $g(x) = g_T(x) = \theta(T/2 - |t|)$.

In the flat model the situation is a bit more tricky. Cosmic time is
now defined by the relation $x^0+x^d = R\, \exp (t/R)$; flat
coordinates cover only half of the de Sitter manifold, namely all
the events such that $x^0+x^d>0$. If we introduce the characteristic
function $h_T(x) = \theta(Re^{T/2R} -x^0 -x^d)\,\theta(x^0 + x^d -
Re^{-T/2R})$ then we have to add the contribution coming from the
other half, i.e. $g(x) = g_T(x)= h_T(x)+ h_T(-x)$. With these
premises we have found that in both models the first factor in
(\ref{1.17}) diverges like $T$; thus it has to be divided by $T$ to
extract a finite result which is the same in both models:
\begin{eqnarray}  \lim_{T\rightarrow \infty}
{\gamma^2 \,C(\kappa)\,\int g(x)\,|F(x)|^2\,dx \over T\int
\ovl{\f_0(x)}\w_{\kappa}(x,\ y)\,\f_0(y)\,dx\,dy}
={\gamma^2\pi\coth(\pi\kappa)^2 \over |\kappa|}\end{eqnarray}

Here the second (unforeseen) result comes in: in contrast to the
Minkowskian case the limiting probability per unit of time  does not
depend on the wavepacket! This result seems to contradict what we
see everyday in laboratory experiments, a well known effect of
special relativity (Eq. \ref{tv}). Furthermore, in contrast with the
violation of particle stability that is exponentially small in the
de Sitter radius, this phenomenon does not depend on how small is
the cosmological constant. How can we solve this paradox and
reconcile the result with everyday experience? The point is that the
idea of probability per unit time (Fermi's golden rule) has no
scale-invariant meaning in de Sitter: if we use the limiting
probability to evaluate amplitudes of processes that take place in a
short time we get a grossly wrong result. This is in strong
disagreement with what happens in the Minkowski case where the
limiting probability is attained almost immediately (i.e. already
for finite $T$). Therefore to describe what we are really doing in a
laboratory we should not take the limit $T\to \infty$ and rather use
the probability per unit of time relative to a laboratory consistent
scale of time. In that case we will recover all the standard wisdom
even in presence of a cosmological constant. But, if an unstable
particle lives a very long time ($>>R$) and we can accumulate observations
then a nonvanishing cosmological constant would radically
modify the Minkowski result and de Sitter invariant result will
emerge. This result should not be shocking: after all erasing any
inhomogeneity is precisely what the quasi de Sitter phase is
supposed to do at the epoch of inflation; in the same way, from the
viewpoint of an accelerating universe all the long-lived particles
look as if they were at rest and so their lifetime would not depend
on their peculiar motion.

 We thank T.~Damour, M.~Gaudin, G.~Gibbons,
M.~Milgram and V.~Pasquier for enlightening discussions. U.~M.
thanks the SPhT  and the IHES for hospitality and support.

\end{document}